\documentclass[prd,nofootinbib,showkeys]{revtex4}
\usepackage{amsmath,graphicx}

\begin{document}

\title{QCD sum rules at finite density in the large-$N_c$ limit: \\
The coupling of the $\rho$-nucleon system to the $D_{13}(1520)$}

\author{Stefan Leupold and Marcus Post}

\thanks{Work supported by DFG.}

\affiliation{Institut f\"ur Theoretische Physik, Justus-Liebig-Universit\"at
Giessen, Germany}

\begin{abstract}
QCD sum rules are studied for the vector-isovector current at finite baryon density in 
the limit of large number of colors $N_c$. For the condensate side it is shown that 
in this limit the four-quark 
condensate factorizes also for the finite density case. At the hadronic side
the medium dependence is expressed in terms of the current-nucleon forward scattering
amplitude. Generalizing vector meson dominance we allow for a direct coupling of
the current to the nucleon as well as a coupling via the $\rho$ meson.
We discuss the $N_c$ dependence of (a) modifications of the pion cloud of the 
$\rho$ meson, (b) mixing with other mesons (in particular $a_1$ and $\omega$)
and (c) resonance-hole
excitations $R\, N^{-1}$. We show that only the last effect survives in the large-$N_c$
limit. Saturating the sum rules with a simple hadronic ansatz which allows for the 
excitation of
the $D_{13}(1520)$ we determine the coupling of the latter to the $\rho$-nucleon
and the photon-nucleon system. These couplings are hard to determine from vacuum 
physics alone.
\end{abstract}

\maketitle

\section{Introduction}
\label{sec:intro}

The question how hadrons once put in a strongly interacting medium change their 
properties provides a very active field of research. In the language of non-perturbative
QCD, in-medium modifications are indicated by the change of condensates like the
quark condensate \cite{Gerber:1989tt} which provides an order parameter of chiral 
symmetry breaking. On the other hand, changes like the melting of the condensates
do not automatically tell what this means for the properties of a particular hadron 
like its mass or lifetime. The situation is such that the condensates are closer
to QCD as the underlying theory of the strong interaction, whereas the hadron 
properties are closer to experimental observation. The QCD sum rule method is
supposed to bridge that gap by connecting integrals over hadronic spectral functions
with an expansion in terms of quark and gluon condensates. Originally they were
introduced for the vacuum \cite{shif79} but later on generalized to in-medium situations
\cite{Bochkarev:1986ex}. 

A particularly interesting probe to study in-medium modifications
are neutral vector mesons. The reason is that such mesons can decay into dileptons.
If such a decay happens in the medium the dileptons leave the system untouched by strong
interactions. In that way in-medium information is carried to the detectors. (For an
overview see e.g.~\cite{Rapp:1999ej}.)
The present paper deals with $\rho$-mesons placed in an infinitely extended
system with finite baryonic density. Such a scenario is an idealization of a finite
nucleus or a heavy-ion collision. For simplicity we study $\rho$-mesons which are at
rest with respect to the nuclear medium.

Originally it was expected that the use of nuclear medium QCD sum rules would yield
model independent predictions for in-medium changes of hadronic properties
--- just like the vacuum sum rules yield in an impressive way parameter-free predictions
of vacuum hadronic properties. 
However, further studies have revealed that for the case of a nuclear medium
there is instead a rather large model dependence
in the possible parameterizations which enter the hadronic side of the sum rules.
In \cite{Hatsuda:1992ez,Hatsuda:1995dy} it has been ``predicted'' that the mass of
the $\rho$-meson should drop in a nuclear medium. The type of parameterization for the 
spectral function of the $\rho$-meson was adopted from the vacuum case: a state with
practically negligible width (in the vacuum this approach is very 
successful \cite{shif79}). On the other hand, in the last years hadronic model builders
have accumulated evidence that in a nuclear medium the $\rho$-meson spectral function 
looks much different from a small-width resonance \cite{herrmann,Chanfray:1993ue,%
Asakawa:1992ht,klingl2,Rapp:1997fs,Peters:1998va,Urban:1998eg,Rapp:1999ej,Post:2000qi,%
Post:2003hu,Bleicher:2000xh,Lutz:2001mi}. 
Clearly, model dependences are inherent in these more complicated in-medium spectral 
functions. In any case it is important to realize that the choice for the 
parameterization of the
spectral information which enters the hadronic side of the sum rules is not an output 
of the sum rule method but an input. The predictions or lessons one deduces from the
sum rules depend on the chosen input. In \cite{Asakawa:1993pq} a specific hadronic model
has been fed into the sum rules. It has been shown that the sum rules require an 
additional downward mass shift on top of the chosen hadronic model. This result was 
basically in agreement with the original stable-$\rho$-meson approximation used in
\cite{Hatsuda:1992ez,Hatsuda:1995dy}. On the other hand, the hadronic model used in
\cite{klingl2} did not yield a sizable mass shift of the $\rho$-meson but a significant
peak broadening. It has been demonstrated that this specific model saturates the sum 
rules without further modifications on top of the hadronic input. 
In \cite{Leupold:1998dg} a one-peak parameterization with arbitrary width and arbitrary
peak position was plugged into the sum rules. This already involved too many free
input parameters (in total four) to deduce predictions. Still a correlation between
width and mass was found which agreed with the previously discussed limiting cases 
of small width and downward mass shift \cite{Hatsuda:1992ez,Hatsuda:1995dy} on the
one hand side and large width and no mass shift \cite{klingl2} on the other.
A parameterization with additional peaks caused by vector--axial-vector mixing
has been studied in \cite{Leupold:2001hj}. On top of these model dependences for the
hadronic input there are also problems on the condensate side. Already for the
vacuum case the size of the four-quark condensate provides a source of uncertainty
(cf.~e.g.~\cite{Leupold:1998dg}). 
The influence of the in-medium behavior of the four-quark condensate has been
studied in detail in \cite{Zschocke:2003hx}.

In \cite{Leupold:1998bt} it was concluded that the nuclear medium sum rule method 
might still 
be useful to constrain a given hadronic model or judge its quality, but
the originally expected predictive power seemed to be lost to a large degree.
It is one purpose of the present work to regain part of this predictive power by 
involving a new QCD based argument.\footnote{Of course, it will turn out that things are
now somewhat more complicated than they were originally thought to be.}
In \cite{'tHooft:1974jz} it has been suggested to treat the number of colors
$N_c$ as a hidden parameter of QCD. In that spirit we will study both sides of the
sum rules as a function of $N_c$. We will demand that both sides match not only for 
$N_c=3$, but for arbitrary $N_c$. In particular, we will be interested in the
large-$N_c$ limit where several important scaling relations are known
\cite{'tHooft:1974jz,witten}. In the large-$N_c$ limit it will be possible to reduce 
the uncertainties on both sides of the sum rules to a large degree. Several hadronic
in-medium effects will drop out in that limit. In addition, the in-medium
four-quark condensate can now be related to the two-quark condensate.

It will turn out that baryonic resonances play an important role on the hadronic side
of the QCD sum rules. Such resonances are formed in collisions of the $\rho$-meson
with a nucleon from the medium. Such a process is also called excitation of a 
resonance-hole state. Especially the $D_{13}(1520)$ resonance might play an important
role for the in-medium properties of the $\rho$-meson at rest 
\cite{Peters:1998va,Post:2000qi,Post:2003hu,Bleicher:2000xh,Cabrera:2000dx}.
It is important to realize, however, that such an analysis relies
on extractions of the resonance parameters from two-pion production data. Unfortunately
such extractions are not completely model-independent, mainly because the $N^*(1520)$
is nominally subthreshold with respect to the $\rho$-nucleon system. Therefore the 
present situation is such that we have to face a rather broad range of possible sizes 
for the coupling of the
$\rho$-nucleon-$D_{13}$ system. As we will see the QCD sum rule method constitutes
a completely independent source of information about this coupling constant.

Concerning the importance of resonances, the present work is close in spirit to 
\cite{Friman:1999wu} where the three-momentum dependence of the sum rules is related
to the excitation of resonances which couple to the $\rho$-nucleon system with a
$p$-wave. Here, however, we study 
$\rho$-mesons at rest and consequently their coupling with nucleons to $s$-wave 
resonances. We also note that $N_c$-scaling arguments were not considered 
in \cite{Friman:1999wu}. 

The paper is organized in the following way: In the next section we study in-medium
QCD sum rules for the $\rho$-meson channel and in particular their condensate side 
in the large-$N_c$ limit. In Sec.~\ref{sec:had} we analyze the $N_c$ scaling of various
hadronic in-medium changes of the hadronic side of the sum rules. In Sec.~\ref{sec:d13}
we become more quantitative and saturate the sum rules by allowing the excitation
of the $D_{13}(1520)$. We determine from the sum rules its couplings to the 
$\rho$-nucleon and photon-nucleon system. Finally we summarize our results in 
Sec.~\ref{sec:sum}.

\section{QCD sum rules at large $N_c$}
\label{sec:cond}

In this work we study the properties of a vector-isovector current 
\begin{equation}
  \label{eq:vecisovec}
j_\mu := \frac12 \left( \bar u \gamma_\mu u - \bar d \gamma_\mu d \right)
\end{equation}
which is at
rest with respect to the nuclear medium. As outlined e.g.~in \cite{Hatsuda:1993bv}
in-medium QCD sum rules can be obtained from an off-shell dispersion relation
which integrates over the energy at fixed (here vanishing) three-momentum of the current.
We also restrict ourselves to small densities $\rho_N$ 
by using the linear density approximation. Effectively this means that the current is at
rest with respect to the nucleon on which it scatters.
The Borel sum rule is given by (cf.~\cite{Leupold:1998dg} and references therein)
\begin{eqnarray}
{1\over \pi M^2} \int\limits^{s_0}_0 \!\! ds \,
{\rm Im} R_{\rm HAD} (s) \, e^{-s/M^2} &=&
{N_c \over 24 \pi^2}
\left(1+{N_c^2 -1 \over N_c} \, {3 \over 8} \,{\alpha_s\over\pi} \right) 
\left( 1 - e^{-s_0/M^2} \right) - {1 \over 4 M^2} \, {\rho_N \over m_N} 
\nonumber \\
&& {}+ {1\over M^4} \, m_q \langle \bar q q\rangle_{\rm med}
+ {1\over 24 M^4} \, \left\langle {\alpha_s \over \pi} G^2 \right\rangle_{\rm med}
+ {1\over 4 M^4} \, m_N a_2 \rho_N 
\nonumber \\
  \label{eq:sumrule}
&& {}-{7 \over 9 M^6} \, {N_c^2 -1 \over N_c^2} \,
\pi\alpha_s \langle {\cal O}^V_4 \rangle_{\rm med}
-{5\over 24 M^6} \, m_N^3 a_4 \rho_N   \,.
\end{eqnarray}

As a first step we will analyze the $N_c$ dependence of the various terms appearing
on the right hand side of the Borel sum rule. The $N_c$ dependence of the perturbative
contribution (first term on the right hand side) is easily obtained from perturbative
QCD. Note that in the spirit of large-$N_c$ QCD \cite{'tHooft:1974jz,witten} the strong 
coupling $\alpha_s$ is $o(1/N_c)$. In total, the perturbative contribution to the
sum rule is $o(N_c)$. 

The continuum threshold $s_0$ is supposed to lie between the
lowest hadronic state in the considered channel --- here the $\rho$-meson --- and the
higher lying states. As the meson masses are $o(N_c^0)$ \cite{witten} 
it is natural to assume that $s_0$ scales in the same way. 

Concerning the external parameter, the density $\rho_N$, we treat it formally as
$o(N_c^0)$. However, this does not matter at all: Below we will compare the terms
linear in the density from both sides of the sum rule with each other only (and not with
the vacuum terms). Hence, in the end the ``size'' of $\rho_N$ in powers of $N_c$
is actually irrelevant.

The next term on the
right hand side of (\ref{eq:sumrule}) is the Landau damping contribution. The nucleon
mass $m_N$ is of order $o(N_c)$ \cite{witten}. 
Therefore the Landau damping term is $o(1/N_c)$,
i.e.~suppressed by two orders of $N_c$ as compared to the leading terms. If all
in-medium contributions scaled in this way, we would not find in-medium changes
in leading order in $N_c$. Indeed it is easy to see that this happens for the case
of finite temperature described by a pion gas \cite{Hatsuda:1993bv}. As we shall see,
however, at finite baryonic density there appear in-medium modifications in leading 
order in $N_c$.

In the linear density approximation the in-medium two-quark condensate
is given by \cite{Drukarev:1991fs,Hatsuda:1992ez}
\begin{equation}
  \label{eq:scal2q}
m_q \langle \bar q q \rangle_{\rm med} = m_q \langle \bar q q \rangle_{\rm vac} 
+ m_q \langle N \vert \bar q q \vert N \rangle \rho_N
= m_q \langle \bar q q \rangle_{\rm vac} + \frac12 \, \sigma_N \rho_N
\end{equation}
where $\vert N\rangle$ denotes a one-nucleon state (with appropriate normalization). 
As the quarks come in $N_c$ colors it is obvious that the vacuum quark condensate 
is $o(N_c)$. But also the in-medium part is of the same order: To see this we
have to show that $\sigma_N = o(N_c)$. In a non-relativistic quark model we can replace
$\bar q q$ by $q^\dagger q$ \cite{Gerber:1989tt}. 
The latter operator counts the number of
constituent quarks. Therefore we obtain\footnote{Note that we neglect effects of isospin
violation.}
\begin{equation}
  \label{eq:sigmanonrel}
{\sigma_N \over m_q} = 2 \langle N \vert \bar q q \vert N \rangle =
\langle N \vert \bar u u + \bar d d \vert N \rangle \approx
\langle N \vert u^\dagger u + d^\dagger d \vert N \rangle = N_c  \,.
\end{equation}
We want to stress here that we do not claim that the non-relativistic quark model
reproduces the right value for $\sigma_N$, but we expect that the scaling with $N_c$
is correctly reproduced. 

The in-medium value of the gluon condensate can be deduced
from the trace anomaly \cite{Drukarev:1991fs,Hatsuda:1992ez}:
\begin{equation}
  \label{eq:gluoncond}
\left\langle {\alpha_s \over \pi} G^2 \right\rangle_{\rm med}
= \left\langle {\alpha_s \over \pi} G^2 \right\rangle_{\rm vac}
- {8 \over {11 \over 3} N_c - {2 \over 3}N_f } m_N^{(0)} \rho_N \,.
\end{equation}
Like the quark condensate the vacuum gluon condensate is $o(N_c)$. The finite
density part, however, is only $o(N_c^0)$, i.e.~suppressed by one power in the number
of colors. Note that the nucleon mass in the chiral limit $m_N^{(0)}$ is $o(N_c)$.
We note in passing that even for $N_c=3$ the in-medium change of the gluon condensate
is rather small \cite{Hatsuda:1992ez}.

Both, the next term and the last one on the right hand side of 
(\ref{eq:sumrule}) come from the nucleon structure functions. 
$a_2$ denotes the fraction of the nucleon momentum carried by the constituent quarks,
i.e.~the first moment of the corresponding structure function.
$a_4$ denotes the corresponding third moment. In the low energy regime which we 
consider ($\mu^2 \approx 1\,$GeV$^2$) most of the momentum is carried by the constituent
quarks. We do not expect that this changes in the large-$N_c$ limit, 
i.e.~$a_2 \approx 1 = o(N_c^0)$. To explore the scaling of $a_4$ we adopt the following
--- obviously over-simplified --- quark model for the description of a nucleon
with momentum $p$: Take $N_c$ constituent quarks each with a mass $m_N/N_c$ and 
momentum $p/N_c$. In this case we get $a_2 = 1$ and $a_4 = 1/N_c^2$. Again we do not
regard that as a realistic model for the structure function but only use it to determine
the $N_c$ scaling behavior. Nonetheless, it is amusing to see that such a model
reproduces the physical values for $a_2$ and $a_4$
within 10\% (cf.~Tab.~\ref{tab:tabnc}). We conclude that both the $a_2$- and the 
$a_4$-term yield in-medium changes of order $N_c$ in (\ref{eq:sumrule}), i.e.~changes
which are as important as the vacuum contributions from quark and gluon condensates.

Finally we turn to the four-quark condensate $\langle {\cal O}^V_4 \rangle_{\rm med}$.
The explicit expression is given e.g.~in \cite{Leupold:2001hj} for three colors. 
In the following we do not need this explicit expression. Instead, we start with
two arbitrary color-neutral two-quark operators $A$ and $B$. 
Note that using Fierz transformations any color-neutral
four-quark operator can 
be written as a sum of products of type $AB$. In the linear density approximation
the in-medium expectation value of $AB$ is given by
\begin{equation}
  \label{eq:lindens}
\langle AB \rangle_{\rm med} = \langle AB \rangle_{\rm vac} 
+ \rho_N \langle N \vert AB \vert N \rangle  \,.
\end{equation}
According to \cite{Novikov:1984jt}
the vacuum part factorizes in the large-$N_c$ limit:
\begin{equation}
  \label{eq:facvac}
\langle AB \rangle_{\rm vac} = \langle A \rangle_{\rm vac} \,\langle B \rangle_{\rm vac} 
+o(N_c)  \,.
\end{equation}
Note that $\langle A \rangle_{\rm vac}$, $\langle B \rangle_{\rm vac} = o(N_c)$.
The nucleon expectation value can be decomposed in the following way:
\begin{equation}
  \label{eq:nuclexp}
\langle N \vert AB \vert N \rangle = 
\langle A \rangle_{\rm vac} \, \langle N \vert B \vert N \rangle 
+ \langle N \vert A \vert N \rangle \, \langle B \rangle_{\rm vac}
+\langle N \vert AB \vert N \rangle_{\rm connected}  \,.
\end{equation}
The last term on the right hand side describes the scattering of $A$ with a nucleon
into $B$ and a nucleon. According to the large-$N_c$ rules developed in \cite{witten}
this contribution is $o(N_c)$. In contrast, the other two terms on the right hand side
of (\ref{eq:nuclexp}) are $o(N_c^2)$ (cf.~(\ref{eq:sigmanonrel})). 
Hence we find (in linear density approximation)
\begin{equation}
  \label{eq:medfac}
\langle AB \rangle_{\rm med} = \langle A \rangle_{\rm vac} \,\langle B \rangle_{\rm vac}
+ \rho_N \,
\left (\langle A \rangle_{\rm vac} \, \langle N \vert B \vert N \rangle 
+ \langle N \vert A \vert N \rangle \, \langle B \rangle_{\rm vac} \right)
+o(N_c)  \,,
\end{equation}
i.e.~''factorization'' of any in-medium four-quark condensate in the large-$N_c$ limit.
The quotation marks are meant to indicate that there is no $\rho_N^2$ term which
would appear if factorization was taken literally.
In particular we find:
\begin{equation}
  \label{eq:faco4}
\langle {\cal O}^V_4 \rangle_{\rm med} 
= \langle \bar q q \rangle_{\rm vac}^2
+ 2 \rho_N \langle N \vert \bar q q \vert N \rangle \langle \bar q q \rangle_{\rm vac}
+ o(N_c)
= \langle \bar q q \rangle_{\rm vac}^2 
+ \rho_N {\sigma_N \over m_q} \, \langle \bar q q \rangle_{\rm vac}
+ o(N_c)  \,.
\end{equation}
Finally the four-quark condensate is multiplied by $\alpha_s = o(1/N_c)$ 
(cf.~(\ref{eq:sumrule})). Hence, in total the ``factorized'' part of the four-quark
condensate enters the sum rule in leading (=linear) order in $N_c$. 

The outcome of our analysis is summarized in Tab.~\ref{tab:tabnc} together with the
actual values we take for the quantitative analysis outlined below. Of course, strictly
speaking we do not know the values of all required quantities for $N_c \to \infty$.
Therefore for the quantitative analysis we use the physical ($N_c =3$) values and
assume that the modifications are not too large.
This is in the same spirit as all other large-$N_c$ approaches. To be formally more
precise we assume the following: for an arbitrary quantity $\cal Q$ which scales like 
$N_c^\gamma$ we assume
\begin{equation}
  \label{eq:scalform}
  \lim_{N_c \to \infty} \left({3^\gamma \over N_c^\gamma} \, {\cal Q} \right) \approx 
\left. {\cal Q} \, \right\vert_{N_c =3} \;.
\end{equation}

Numerically we find that most of the in-medium changes of the right hand side of 
(\ref{eq:sumrule}) found for $N_c=3$ remain present in the large-$N_c$ limit. 
\begin{table}[htbp]
  \centering
  \begin{tabular}{|c|c|c|c|}
\hline
quantity & size & scaling & ref. \\ \hline \hline
$g_V$ & 6.05 & $1/\sqrt{N_c}$ & \cite{klingl2} \\ \hline
$m_V$ & $770\,$MeV & 1 & \cite{pdg02} \\ \hline
$s_0$ & $1.3\,$GeV$^2$ & 1 & \cite{klingl3,Marco:1999xz,Leupold:2003zb} \\ \hline
$m_N$ & $940\,$MeV & $N_c$ & \cite{pdg02} \\ \hline
$\sigma_N$ & $45\,$MeV & $N_c$ & \cite{Hatsuda:1992ez} \\ \hline
$a_2$ & 0.9 & 1 & \cite{Hatsuda:1992ez} \\ \hline
$a_4$ & 0.12 & $1/N_c^2$ & \cite{Hatsuda:1992ez} \\ \hline
$\alpha_s$ & 0.36 & $1/N_c$ & \cite{Hatsuda:1992ez} \\ \hline
$\langle \bar q q \rangle_{\rm vac}$ & $(-240 \,{\rm MeV})^3$ & $N_c$ &
\cite{GOR,Leupold:2003zb} \\ \hline
$m_q$ & $6 \,$MeV & 1 & \cite{pdg02} \\ \hline
$m_{D13}$ & $1520\,$MeV & $N_c$ & \cite{pdg02} \\ \hline
$m_{D13}-m_N$ & $580\,$MeV & $1$ & \cite{pdg02} \\ \hline
$f_\pi$ & $92\,$MeV & $\sqrt{N_c}$ & \cite{pdg02} \\ \hline
  \end{tabular}
  \caption{Sizes and $N_c$-scaling of all relevant quantities.}
  \label{tab:tabnc}
\end{table}

To be more sensitive to the in-medium modifications we differentiate the 
Borel sum rule (\ref{eq:sumrule}) with respect to the density:
\begin{eqnarray}
{1 \over \pi M^2} \int\limits_0^{s_0} \!\! ds \, e^{-s/M^2}
\left. {\partial \over \partial \rho_N} {\rm Im}R_{\rm HAD}(s,\rho_N) 
\right\vert_{\rho_N =0} 
= {1 \over M^2} {N_c \over 24 \pi^2}
\left(1+{N_c^2 -1 \over N_c} \, {3 \over 8} \,{\alpha_s\over\pi} \right) e^{-s_0/M^2} \,
s_0' + {1\over M^4} \,c_2
+ {1 \over M^6} \, c_3
  \label{eq:bsr}
\end{eqnarray}
with
\begin{subequations}
\begin{eqnarray}
  \label{eq:defc2}
  c_2 & = & {m_N a_2 \over 4} + {\sigma_N \over 2} \,,
\\[0.5em]
  \label{eq:defc3}
  c_3 & = & -{7 \over 9} \, {N_c^2 -1 \over N_c^2} \,\pi \alpha_s 
\langle \bar q q \rangle_{\rm vac}
{\sigma_N \over m_q} - {5 \over 24} \, m_N^3 a_4  \,,
\end{eqnarray}
\end{subequations}
$s_0 = s_0(\rho_N = 0)$ and
\begin{equation}
  \label{eq:s0der}
s_0'= \left. {d s_0 \over d \rho_N} \right\vert_{\rho_N =0}  \,.
\end{equation}
Note that we have only kept the terms which remain present in the large-$N_c$ limit,
i.e.~we have neglected the Landau damping contribution and the in-medium change of the
gluon condensate. Next we rewrite (\ref{eq:bsr}):
\begin{eqnarray}
{1 \over \pi} \int\limits_0^{s_0} \!\! ds \, e^{(s_0-s)/M^2}
\left. {\partial \over \partial \rho_N} {\rm Im}R_{\rm HAD}(s,\rho_N)
\right\vert_{\rho_N =0}
= {N_c \over 24 \pi^2}
\left(1+{N_c^2 -1 \over N_c} \, {3 \over 8} \,{\alpha_s\over\pi} \right) \,
s_0' + {1\over M^2} \,c_2 \, e^{s_0/M^2} 
+ {1 \over M^4} \, c_3 \, e^{s_0/M^2} \,,
  \label{eq:bsr2}
\end{eqnarray}
expand both sides in powers of $1/M^2$ and compare the corresponding coefficients on
right and left hand side:
\begin{subequations}
\label{eq:FESR}
  \begin{eqnarray}
    \label{eq:wFESR0}
{1 \over \pi} \int\limits_0^{s_0} \!\! ds \, 
\left. {\partial \over \partial \rho_N} {\rm Im}R_{\rm HAD}(s,\rho_N)
\right\vert_{\rho_N =0} 
& = & {N_c \over 24 \pi^2}
\left(1+{N_c^2 -1 \over N_c} \, {3 \over 8} \,{\alpha_s\over\pi} \right) \, s_0'  \,,
\\
    \label{eq:wFESR1}
{1 \over \pi} \int\limits_0^{s_0} \!\! ds \, (s_0-s)
\left. {\partial \over \partial \rho_N} {\rm Im}R_{\rm HAD}(s,\rho_N)
\right\vert_{\rho_N =0} 
& = & c_2  \,,
\\
    \label{eq:wFESR2}
{1 \over \pi} \int\limits_0^{s_0} \!\! ds \, (s_0-s)^2 
\left. {\partial \over \partial \rho_N} {\rm Im}R_{\rm HAD}(s,\rho_N)
\right\vert_{\rho_N =0} 
& = & 2 \, (c_2 \, s_0 + c_3)  \,.
  \end{eqnarray}
\end{subequations}
In this way we have obtained weighted finite energy sum rules. The advantage of
finite energy type sum rules as compared to Borel sum rules lies in the fact that
with the former we have got rid off the Borel mass and the problem how to determine
a reliable Borel window etc. (see e.g.~\cite{Leupold:1998dg} and references therein). 
On the other hand, the standard finite energy sum rules are rather sensitive to
the modeling of the transition region from the hadronic part ${\rm Im}R_{\rm HAD}$
to the continuum (see e.g.~\cite{Leupold:2001hj} and references therein). Indeed,
the first equation (\ref{eq:wFESR0}) is plagued by that problem. The latter two
equations, however, are not since the transition region is suppressed by powers of
$(s_0-s)$ \cite{maltman}. Therefore (\ref{eq:wFESR1}) and (\ref{eq:wFESR2}) are
more reliable as they are insensitive to details of the
threshold modeling at $s_0$. Hence these weighted finite energy sum rules combine the
advantages of Borel and standard finite energy sum rules. In general, the disadvantage 
is that there are only two properly weighted finite energy sum rules as compared to 
three standard finite energy sum rules. In our case, however, this does not reduce
the available information: The in-medium change of the threshold parameter encoded in
$s_0'$ is anyway unknown {\it a priori}. Fortunately, it only appears in the first 
(anyway less reliable) sum rule (\ref{eq:wFESR0}). The two preferable sum rules
(\ref{eq:wFESR1}) and (\ref{eq:wFESR2}) are independent of $s_0'$. 
We shall use them for the subsequent analysis. Note that the vacuum threshold $s_0$
appears in (\ref{eq:wFESR1}) and (\ref{eq:wFESR2}). This, however, can be fixed by
an independent vacuum sum rule analysis which is free of all in-medium uncertainties.
For the actual calculation we adopt the point of view of 
\cite{klingl3,Marco:1999xz,Leupold:2003zb} and use 
$s_0 \approx (3/N_c)\,(4 \pi f_\pi)^2 \approx 1.3\,$GeV$^2$ with the pion decay constant
$f_\pi$. Note that the latter scales with $\sqrt{N_c}$ as can be deduced e.g.~from the 
Gell-Mann--Oakes--Renner relation \cite{GOR}.

\section{Hadronic in-medium changes at large $N_c$}
\label{sec:had}

We now turn to the left hand side of the sum rules (\ref{eq:sumrule}) or 
(\ref{eq:FESR}). It is
well-known that the vector-isovector current $j_\mu$ strongly couples to the 
$\rho$-meson. In the vector meson dominance (VMD) picture which is phenomenologically
rather successful it is even assumed that
all the interaction of $j_\mu$ with hadrons is mediated by the $\rho$-meson 
\cite{sakuraiVMD}. 
In the following we will not adopt the strict VMD picture. Nonetheless,
the in-medium modifications of the $\rho$-meson will significantly influence the
current $j_\mu$. Suppose for simplicity that in the medium there is one additional
channel besides the $\rho$-meson which couples directly to the current. 
Later on this channel will
be specified to be a resonance-hole excitation. For the moment, however, we keep the
formalism a little bit more general. The vertex which gives the strength of the
coupling of the current to the considered channel is given
by $f_\gamma$. The channel also couples to the $\rho$-meson with strength $f_\rho$. 
A generalization to several channels is straightforward.
Following \cite{Friman:1997tc} we get
\begin{equation}
  \label{eq:imr}
{\rm Im}R_{\rm HAD}(s,\rho_N) = -{1 \over g_V^2 s} \left[
{\rm Im}\Pi_V(s) \vert d_V(s)-1 \vert^2 + {\rm Im}\Pi_B(s) \vert d_V(s) -r \vert^2
\right]
\end{equation}
with
\begin{equation}
  \label{eq:dvdef}
d_V(s) = {s-r\Pi_B(s)-\Pi_V(s)  \over s-m_V^2-\Pi_B(s)-\Pi_V(s)}  \,,
\end{equation}
\begin{equation}
  \label{eq:defr}
  r = {f_\gamma \over f_\rho}
\end{equation}
and in the linear density approximation
\begin{equation}
  \label{eq:defpib}
\Pi_B(s) = \rho_N T(s)
\end{equation}
with the $\rho$-$N$ forward scattering amplitude $T$. The vacuum self energy of the
$\rho$-meson is given by $\Pi_V$, $m_V$ denotes the vacuum $\rho$-meson mass and $g_V$
the $\rho$-pion-pion coupling.

\begin{figure}[htbp]
  \centering
    \includegraphics[keepaspectratio,width=0.7\textwidth]{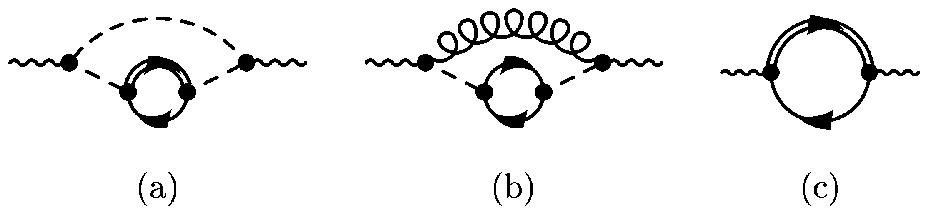}
  \caption{Typical Feynman diagrams for various in-medium modifications. The wavy lines
denote $\rho$-mesons, the dashed lines pions, the full lines nucleons, the double
lines baryonic resonances and nucleons and the curly line $\omega$- and $a_1$-mesons.
See main text for details.}
  \label{fig:diag}
\end{figure}

In a nuclear
medium there are various ways how the nucleons which form the medium can interact with
the $\rho$-meson or directly with the current $j_\mu$. We classify the important
in-medium effects in the following way:
\begin{itemize}
\item[(a)] Modifications of the pion cloud (cf.~Fig.~\ref{fig:diag}a)
\cite{herrmann,Chanfray:1993ue,Asakawa:1992ht,Asakawa:1993pq,klingl2,%
Rapp:1997fs,Urban:1998eg,Rapp:1999ej,Cabrera:2000dx}:
In the vacuum the 
$\rho$-meson can decay into pions. The latter get strongly modified in the medium,
e.g.~by their coupling to $\Delta(1232)$-hole states. 
This in turn changes the $\rho$-meson
properties. 
\item[(b)] Mixing with other mesons (cf.~Fig.~\ref{fig:diag}b):
The nucleons which form the medium
carry a pion cloud. These pions can interact with the $\rho$-mesons (or directly with
the current) and cause a mixing with other mesons, e.g.~with the $\omega$-meson
\cite{klingl2} or with the $a_1$-meson (chiral mixing) 
\cite{Chanfray:1998ws,Krippa:1998ss,Krippa:2000jh,Kim:1999pb,Leupold:2001hj}.
\item[(c)] Excitation of resonance-hole pairs (cf.~Fig.~\ref{fig:diag}c):
Instead of these indirect interactions via pions the $\rho$-meson can
also couple directly to a nucleon from the medium 
and form a nucleon or a baryonic resonance. For $\rho$-mesons at rest 
which we study here,
one can only form baryons which couple to the $\rho$-nucleon system in an $s$-wave
\cite{Peters:1998va,Post:2000qi,Post:2003hu,Rapp:1998ei,Kim:1999pb,%
Bleicher:2000xh,Cabrera:2000dx}.
This excludes e.g.~the nucleon and the $\Delta(1232)$ state. Concerning moving
$\rho$-mesons and their $p$-wave interaction 
cf.~also e.g.~\cite{Friman:1997tc,Rapp:1997fs,Friman:1999wu}.
\item[(d)] Changes of the vacuum structure at the underlying quark-gluon level 
(described e.g.~by Brown-Rho scaling \cite{Brown:1991kk}): 
If the density and/or temperature
of the medium is high enough one expects a transition to a state where the single
quarks (and gluons) are the relevant degrees of freedom \cite{Hwa:1990xg,Hwa:1995wt}.
As a precursor of these changes
the non-perturbative QCD vacuum structure already changes in the hadronic medium
(expressed e.g.~in the melting of the condensates \cite{Gerber:1989tt,Drukarev:1991fs}).
The in-medium change of the underlying vacuum structure influences the quark properties 
and in turn the properties of the hadrons which consist out of quarks. 
From the hadronic point of view it is not clear whether such effects are distinct
\cite{Asakawa:1992ht,Kim:1999pb}
from the hadronic effects discussed above or whether it is just a different language
for the same physics. 
It is also not clear whether these ``exotic'' effects show up at all in linear 
order in the density. After all, the
linear density approximation describes the interaction of the probe (the current) with
one nucleon at a time. The key ingredient is the current-nucleon forward scattering
amplitude, i.e.~a vacuum quantity. 
In the following we will disregard such ``exotic'' effects and figure out whether the
sum rule can be saturated by the standard hadronic interactions discussed above. 
\end{itemize}
Above we have described in-medium modifications of the $\rho$-meson.
Note, however, that all these effects (a-d) 
can also directly influence the current instead of the $\rho$-meson.

As a next step we will analyze the $N_c$ scaling behavior of the hadronic effects (a-c)
discussed above. We recall the large-$N_c$ counting rules developed in 
\cite{'tHooft:1974jz,witten}:
\begin{itemize}
\item Meson mass: $o(N_c^0)$.
\item Baryon mass: $o(N_c)$.
\item Mesonic interactions are suppressed in the large-$N_c$ limit. The decay width
of a meson into two other mesons is e.g.~$o(1/N_c)$. Therefore, in vacuum mesons are
stable in the large-$N_c$ limit. 
\item In contrast, meson-baryon interactions are not suppressed. Meson-baryon 
scattering amplitudes are $o(N_c^0)$. 
A meson-baryon-baryon vertex might even be enhanced,
$o(\sqrt{N_c})$. Therefore, in general, baryonic contributions e.g.~to
the meson self energy are as important as the meson mass. 
This actually opens the possibility
that a baryonic medium influences the meson properties in leading order in $N_c$.
\end{itemize}
There are remarkable exceptions to the last rule: The pseudovector interaction of a pion
with a nucleon is proportional to the inverse of the pion decay constant $f_\pi$. 
Recall that the latter scales with $\sqrt{N_c}$. 
Therefore the pion-nucleon-nucleon vertex is suppressed in the
large-$N_c$ limit. Also the vector interaction of the $\rho$-meson with the nucleon
is suppressed (whereas the tensor interaction is not) 
\cite{Mattis:1988hg,Mattis:1989hf,Mattis:1989nh,Donohue:1989fe}. This fits well
with the approach to introduce the coupling of the $\rho$-meson to other hadrons
via minimal substitution: The $\rho$-pion-pion coupling constant $g_V$ is
$o(1/\sqrt{N_c})$. Minimal substitution introduces the same coupling for the 
$\rho$-nucleon-nucleon vertex and
yields the suppression mentioned above. In addition, the principle of minimal 
substitution creates an elementary four-point coupling (Kroll-Ruderman type%
)
$\rho$-nucleon-pion-nucleon from the pseudovector pion-nucleon interaction. 
Consequently the corresponding inelastic scattering amplitude $\rho$-nucleon to
pion-nucleon is $o(1/N_c)$ and not $o(N_c^0)$. 

It is also interesting to realize that the large-$N_c$ limit yields a justification
for two commonly used approximations in the vacuum QCD sum rule approach, namely
factorizing the four-quark condensate and approximating the $\rho$-meson peak by
a delta function \cite{shif79}. Hence the standard treatment of the $\rho$-meson in 
the QCD sum rule method is practically identical to its large-$N_c$ treatment.
In turn one can therefore conclude that one gets reasonable results for the vacuum
properties of the $\rho$-meson by combining QCD sum rules with large-$N_c$ 
considerations. In the present work we apply that combination of techniques to the
in-medium situation.

Let us now come back to the hadronic in-medium effects (a-c) discussed above. Both
effects (a) and (b) are mediated by pions. The nucleon interaction to the pions 
already generates a suppression due to the $1/f_\pi$ factors. In addition, 
the coupling of the pions to the $\rho$-meson is suppressed.
In total, the effects (a) and (b) are suppressed in the large-$N_c$ approximation
as compared to the vacuum properties. This can also be deduced in detail by
inspecting the appropriate expressions given e.g.~in \cite{klingl2}.
In contrast, the direct coupling of the $\rho$-meson to nucleons discussed in (c)
in general is not suppressed. Therefore, the large-$N_c$ approximation allows to 
disentangle different hadronic in-medium effects. Studying both
sides of the sum rule as functions of $N_c$ suggests that the sizable in-medium
modification of the condensate side does not reflect modifications of
the pion cloud or mixing with other mesons but the excitation of resonance-hole states.
This is a simple consequence of the fact that the effects (a) and (b) show a scaling
with $N_c$ which does not comply with the condensate side. In turn, we can now
use the sum rule in the large-$N_c$ approximation to quantify the effect of
resonance-hole states. 

So far we have tried to be as model-independent as possible
concerning the modeling of hadronic interactions. To make further progress, however,
we need some phenomenological input. In \cite{Peters:1998va,Post:2000qi,Post:2003hu}
the influence of baryonic resonances on the in-medium properties of the $\rho$-meson
were thoroughly discussed. It is shown there that the most important in-medium change
for $\rho$-mesons at rest is caused by the excitation of the $D_{13}(1520)$ resonance
(see also \cite{Bleicher:2000xh}). We should add, however, that such an analysis relies
on extractions of the resonance parameters from two-pion production data. Unfortunately
such extractions are not completely model-independent, mainly because the $N^*(1520)$
is nominally subthreshold with respect to the $\rho$-nucleon system. Therefore different
analyses yield a rather broad range of possible sizes for the coupling of the
$\rho$-nucleon-$D_{13}$ system. For a detailed discussion we refer to \cite{Post:2003hu}.
It is one purpose of the present work to determine that coupling constant by a completely
independent approach, the QCD sum rule method. In the following we shall neglect
all other baryonic resonances besides the $D_{13}(1520)$. Note again that several other
possibly important channels like the nucleon or the $\Delta(1232)$ are $p$-waves.
Consequently they do not couple to a $\rho$-meson which is at rest relative to the 
nucleon from the medium. We will try to saturate the sum rules (\ref{eq:FESR})
by including only the $D_{13}(1520)$ in the $\rho$-nucleon scattering amplitude
$T$ introduced in (\ref{eq:defpib}).

\section{The $\rho$-nucleon-$D_{13}$ system}
\label{sec:d13}

To utilize the finite energy sum rules (\ref{eq:FESR}) we need the density derivative
of (\ref{eq:imr}). In the large-$N_c$ limit, $\Pi_V \to 0$ and
\begin{eqnarray}
\left. {\partial \over \partial \rho_N} {\rm Im}R_{\rm HAD}(s,\rho_N)
\right\vert_{\rho_N =0}
& = & -{1 \over g_V^2 s} \left\{
{\rm Im}T(s) \left[
(r-1)^2-m_V^4 {d \over ds} {\rm Re}D_V(s) - 2 m_V^2 \, (r-1) \, {\rm Re}D_V(s) \right]
\right.  \nonumber \\  && \left. \phantom{mmmm}
+ {\rm Re}T(s) \left[
-m_V^4 {d \over ds} {\rm Im}D_V(s) - 2 m_V^2 \, (r-1) \, {\rm Im}D_V(s)
\right] \right\}
  \label{eq:imrder}
\end{eqnarray}
with the free $\rho$ meson propagator
\begin{equation}
  \label{eq:deffreerhoprop}
D_V(s) = {1 \over s - m_V^2 + i \epsilon}  \,.
\end{equation}
Allowing for the excitation of a $D_{13}$ state the imaginary part of the
$\rho$-$N$ forward scattering amplitude for vanishing three-momentum is given by
\begin{equation}
  \label{eq:scattamp}
{\rm Im}T(s) = \left( {f_\rho \over m_V} \right)^2 \, {2 \over 3} \, s \,
{\rm Im}D_{D13}(q_0=\sqrt{s}+m_N,\vec q =0)
\end{equation}
with the propagator $D_{D13}$ to be specified below. The amplitude is
deduced from the non-relativistic lagrangian \cite{Post:2003hu}
\begin{equation}
  \label{eq:lagrd13}
{\cal L} = {f_\rho \over m_V} \, \psi_R^\dagger S_i^+ \tau_a \psi_N \,
(\partial_i \rho_0^a - \partial_0 \rho_i^a) + \mbox{h.c.}
\end{equation}
where $\psi_R$ denotes the $D_{13}$-resonance isospinor-spinor,
$\psi_N$ the nucleon isospinor-spinor, $\rho_\mu^a$ the $\rho$-meson isovector-vector,
$S_i^+$ the spin-$\frac12$ to $\frac32$ transition operator and $\tau_a$ the (Pauli)
isospin matrix. Finally h.c.~stands for hermitian conjugate.

The simplest version of a $D_{13}$ propagator is a non-relativistic one 
with vanishing width
\begin{equation}
  \label{eq:d13prop}
  D_{D13}(q_0,\vec q) = {1 \over q_0 - m_{D13}-{\vec q^2 \over 2 m_{D13}} + i \epsilon}
\,.
\end{equation}
Here $m_{D13}$ is the resonance mass. Following \cite{coleman-erice} we assume that the
mass difference between the nucleon and its excited state is $o(N_c^0)$.
Whether neglecting the width of the baryon resonance is a reasonable approximation
will be discussed later. For the moment we are aiming at an expression with a minimal
number of free parameters. For the imaginary part of the forward scattering amplitude 
we obtain
\begin{equation}
  \label{eq:d13im}
{\rm Im}T(s) = -\left( {f_\rho \over m_V} \right)^2 \, {2 \over 3} \, 2\pi \, 
(m_{D13}-m_N)^3 \, \delta\left(s-(m_{D13}-m_N)^2 \right) \,.
\end{equation}
We determine the corresponding real part from a dispersion relation such that the 
amplitude does not diverge for large $s$:
\begin{equation}
  \label{eq:disp1}
{\rm Re}T(s) = - {1 \over \pi} \int\limits_0^{s_0} \!\! ds' \, 
{{\rm Im}T(s') \over s-s'} + \mbox{const.}
\end{equation}
and determine the constant such that the real part vanishes at the photon point 
(gauge invariance)
\begin{equation}
  \label{eq:disp2}
  {\rm Re}T(s) = - {1 \over \pi} \int\limits_0^{s_0} \!\! ds' \, 
{\rm Im}T(s') \left( {1 \over s-s'} + {1 \over s'} \right)  \,.
\end{equation}
Thus
\begin{eqnarray}
  \label{eq:d13re}
{\rm Re}T(s) & = & \left( {f_\rho \over m_V} \right)^2 \, {2 \over 3} \, 
2 \, (m_{D13}-m_N) {s \over s-(m_{D13}-m_N)^2} 
\nonumber \\ 
& = & \left( {f_\rho \over m_V} \right)^2 \, {2 \over 3} \, s
\left[
{1 \over \sqrt s - (m_{D13}-m_N)} - {1 \over \sqrt s + (m_{D13}-m_N)}
\right]  \,.
\end{eqnarray}
The last expression shows that we have achieved the inclusion of an $s$- and a
$u$-channel process.

\begin{figure}[htbp]
  \centering
    \includegraphics[keepaspectratio,width=0.7\textwidth]{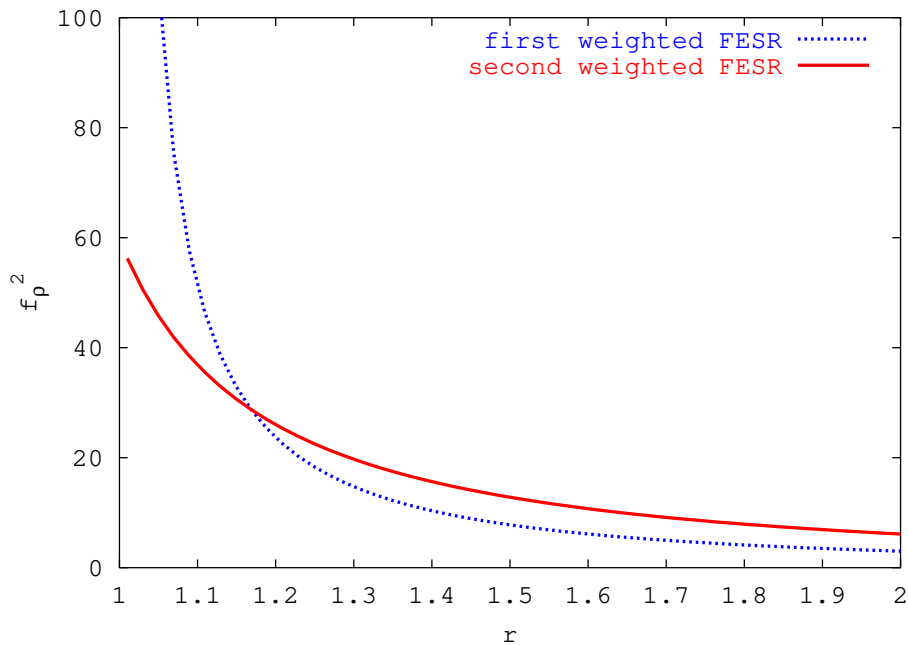}
  \caption{The weighted finite energy sum rules (FESR) given in (\ref{eq:wFESR1}) and
(\ref{eq:wFESR2}) relate the ratio $r$ defined in (\ref{eq:defr}) with the coupling
(squared) $f_\rho^2$ from (\ref{eq:lagrd13}). The outcome of 
(\ref{eq:wFESR1})/(\ref{eq:wFESR2}) is given by the dotted/full line.
See main text for details.}
  \label{fig:wfesr}
\end{figure}

Now we are in the position to determine the free parameters $f_\rho$ and $r$ from
the sum rules (\ref{eq:wFESR1}) and (\ref{eq:wFESR2}) using (\ref{eq:imrder}),
(\ref{eq:d13im}) and (\ref{eq:d13re}). Both of the two sum rules relates $f_\rho^2$
to $r$. These relations are depicted in Fig.~\ref{fig:wfesr}.
The two relations intersect for
\begin{equation}
  \label{eq:fres}
f_\rho^2 \approx 28.7 \,, \qquad 
r \approx 1.17
\end{equation}
which finally yields
\begin{equation}
  \label{eq:fgamres}
\vert f_\gamma \vert = \vert r \, f_\rho \vert \approx 6.26  \,.
\end{equation}

Before we compare these results with other approaches we take these values 
(\ref{eq:fres}) as an input for (\ref{eq:wFESR0}) and determine the in-medium
change of the threshold parameter expressed by $s_0'$. We find
\begin{equation}
  \label{eq:ders0res}
  s_0' \approx 0.002 \, {s_0 \over \rho_{\rm n.m.}}
\end{equation}
with normal nuclear matter density
\begin{equation}
  \label{eq:nuclmatt}
\rho_{\rm n.m.} = 0.17 \,{\rm fm}^{-3}\,.
\end{equation}
Recall that the sum rule (\ref{eq:wFESR0}) is less reliable as already discussed above.
Still we can take (\ref{eq:ders0res}) as an indication that the change of the
threshold parameter is extremely small. This finding is in contrast to the traditional
result that the threshold sizably drops in a nuclear medium 
(e.g.\cite{Hatsuda:1992ez,klingl2,Leupold:1998dg}). Both results are easy to interpret:
The $\rho$-meson is the lowest state
in the vector-isovector channel. The higher states are effectively taken into account
by the continuum contribution which starts at $s_0$. 
Traditionally a dropping mass of the $\rho$-meson is deduced from the QCD sum rule 
approach (cf.~discussion in the introduction). It is natural to assume in such a scenario
that also the masses of the higher-lying states drop which effectively leads to a 
lowering of the threshold $s_0$. On the other hand, the present picture deduced from
the large-$N_c$ considerations suggests that there is no change of the $\rho$-peak
(besides level repulsion, see below), but the appearance of an additional peak, the
collective resonance-hole excitation. If the $\rho$-peak is unchanged there is
no reason why the higher-lying states should move. Consequently the threshold remains
more or less the same.

It is also interesting to discuss the pole structure of (\ref{eq:imr}) or
(\ref{eq:dvdef}), respectively. Besides the $\rho$-meson pole which is already
present in the vacuum, there appears a second pole caused by the collective
resonance-hole excitation. At normal nuclear matter density the two poles are at
$m_1 \approx (1480-940)\,$MeV = $540\,$MeV and $m_2 \approx 830\,$MeV. As compared
to the ``free'' poles $(1520-940)\,$MeV = $580\,$MeV and $770\,$MeV, level repulsion
shifts the collective pole downwards and the $\rho$-pole upwards.
Note that we went beyond the linear density approximation by putting the 
self energy (\ref{eq:defpib}) in the denominator of (\ref{eq:dvdef}). Only by this
iteration the $\rho$-peak shifts.

Our final numerical result is given by (\ref{eq:fres}). Nonetheless it is illuminating 
to obtain an analytic result which provides a rough estimate. For that purpose we observe
that our final result for $r$ is rather close to 1. Therefore we evaluate the weighted
finite energy sum rules for $r=1$ (which is strict VMD \cite{Friman:1997tc}). 
The left hand side of (\ref{eq:wFESR1}) vanishes --- if the coupling does not
diverge. (This can also be observed in the
strong rise of the dotted curve in Fig.~\ref{fig:wfesr}.) Therefore it is of no use
for our present purpose. In contrast, (\ref{eq:wFESR2}) does not vanish. Inspecting
the right hand side, we observe that the numerically strongest contribution comes from
the $a_2$ term (cf.~(\ref{eq:defc2}), (\ref{eq:defc3}) and Tab.~\ref{tab:tabnc}).
This is actually an interesting observation since the size of this term is rather save
as compared to the size of $\sigma_N$. The latter enters the in-medium parts of the
two- and four-quark condensate. Therefore we feel rather comfortable with our approach
where the strongest contribution comes from a well-known term.
Using for simplicity $a_2 \approx 1$ and 
$g_V^2 s_0/m_V^2 \approx 24 \pi^2 /N_c$ \cite{Leupold:2003zb} and neglecting all other
terms we obtain after some calculation:
\begin{equation}
  \label{eq:rough}
f_\rho^2 \approx {9  \pi^2 m_N \over (m_{D13}-m_N) \, N_c} \approx 48\,, \quad \quad 
\mbox{rough estimate!}
\end{equation}
This indicates that parametrically $f_\rho$ and therefore also the $\rho$-nucleon 
scattering amplitude is $o(N_c^0)$ as it should be.

Clearly the value for the $\rho$-nucleon-$D_{13}$ coupling constant crucially depends on 
the input lagrangian used for the calculation.
Therefore for a comparison of our results to other approaches it is more 
illuminating to calculate
quantities which are measurable or at least closer to measurable ones. For that purpose
we use $f_\rho$ as an input for eq.~(9) in \cite{Post:2003hu} to calculate the partial 
decay width $D_{13} \to \rho N$. We obtain $\approx 10\,$MeV. This value compares 
favorably with $12\,$MeV obtained by \cite{Vrana:1999nt}. It is smaller than 
$26\,$MeV deduced in \cite{Manley:1984jz,Manley:1992yb} but much bigger than 
$2\,$MeV from \cite{Lutz:2001mi}. Note that the $N^*(1520)$ is nominally subthreshold 
with respect to the decay products $\rho$-meson plus nucleon. The decay is only possible
due to the finite width of the $\rho$-meson. This has two interrelated aspects:
Technically, for a reasonable width 
calculation for such a baryonic resonance it is crucial to use a proper width 
(self-energy)
parameterization for the $\rho$-meson. Therefore we use the momentum-dependent width
caused by the two-pion decay of the $\rho$-meson \cite{Post:2003hu}. The second
aspect is the principal one already discussed above: Extracting a coupling constant
for such a decay situation from the two-pion data introduces a rather high model
dependence into the analysis. It is appealing that we have extracted here additional
information from a completely different source.

We can also calculate the partial decay width of the $D_{13}$ into a nucleon and an
isovector photon:
\begin{equation}
  \label{eq:decphoton}
\Gamma_{N\gamma} = {\mu^2 m_N q^3_{cm} \over 3 \pi m_{D13}}
\end{equation}
with the center-of-mass momentum
\begin{equation}
  \label{eq:cmdef}
q_{cm} = {m_{D13}^2 - m_N^2 \over 2 m_{D13}}
\end{equation}
and \cite{Friman:1997tc}
\begin{equation}
  \label{eq:defmufp}
\mu = {e f_\gamma \over g_V m_V}
\end{equation}
where $e$ denotes the electric charge. We obtain $\approx 1\,$MeV which can be compared
to the analysis of Arndt et al.~\cite{Arndt:1996ak} where $\approx 0.51\,$MeV is deduced.

At first sight, one might feel disappointed about this factor of two mismatch.
However, we do not share that point of view for the following reasons:
First, it is important to stress that we have presented a completely parameter-free 
determination of the coupling constants. It is by far non-trivial to obtain agreement
within a factor of two. Second, one has to recall that the uncertainty in 
the value for the decay width into $\rho$-nucleon is
much larger than a factor of two, as discussed above. Therefore we could actually
narrow the range of possible values in that case.
Third, several approximations entered the calculation:
We worked in the large-$N_c$ limit (otherwise we could not sort out the various
hadronic in-medium effects). We neglected all other resonances besides the $D_{13}$
(otherwise we would not have enough equations to determine the various coupling 
constants). We used a non-relativistic model for $D_{13}$ and neglected the width of 
the $D_{13}$. It is interesting to note that this last point is actually not crucial:
We have found numerically that the obtained values 
for the coupling constants $f_\rho$ and $f_\gamma$ practically do not depend on the 
value for the $D_{13}$ mass. Therefore a finite width cannot change the results.
In view of all these considerations we regard our results for the coupling constants
as rather satisfying.

Before summarizing 
we would like to comment on the possible nature of the $D_{13}(1520)$-resonance.
It has been suggested recently \cite{Lutz:2001mi,Kolomeitsev:2003kt} that this 
resonance (among others) could be a consequence of coupled-channel dynamics and not
a three-quark resonance. In the present work we have treated the resonance as an 
elementary field (cf.~(\ref{eq:lagrd13})). This, however, merely tells about the way how
to effectively describe the resonance and not about its nature. The more relevant
question is whether the resonance survives in the large-$N_c$ limit. Clearly a 
three-quark resonance does. Whether this is also true for a dynamically created resonance
actually depends on the details of the underlying model. In particular, it depends on the
question whether the four-point meson-baryon interactions studied in 
\cite{Lutz:2001mi,Kolomeitsev:2003kt} survive or become suppressed in the 
large-$N_c$ limit. In principle, such vertices can be $o(N_c^0)$ as discussed above.
In this case our analysis still applies. In \cite{Lutz:2001mi} no statement about
the $N_c$ scaling of the coupling constants was made and could be made, as they were
introduced on a phenomenological level. In contrast, in \cite{Kolomeitsev:2003kt} it
was claimed that the main mechanism for dynamical resonance formation is the
Weinberg-Tomozawa interaction. The latter is completely fixed by the pion decay constant
and produces four-point vertices of $o(1/N_c)$. For such a scenario our analysis does
not fit, since no resonance would be formed in the large-$N_c$ limit. 
Nonetheless the analysis of \cite{Kolomeitsev:2003kt} is not and was not
meant to be a complete analysis of the negative parity, spin-$3/2$ resonances.
Only channels with Goldstone boson and baryon decuplet states where considered there
while e.g.~channels with vector mesons and baryon octet states where missing.
It is not clear whether the interaction in the latter channels (and its mixing with the
former ones) also vanishes in the
large-$N_c$ limit. {\it A priori} there is no reason why it should vanish. If this
interaction was sufficient to form the $D_{13}(1520)$ --- maybe with a different mass
--- our analysis might still be applicable.

\section{Summary}
\label{sec:sum}

The present work relies on two basic assumptions. First, that the QCD sum rule
approach provides a connection between hadronic and quark-gluon properties
not only for $N_c=3$ but for arbitrary number of colors, in particular for 
$N_c \to\infty$. Second, that for $N_c \to \infty$ the input parameters 
(Tab.~\ref{tab:tabnc}) do not differ much from their $N_c =3$ values (besides the
appropriate rescaling factors $3/N_c$ to some power). The condensate side of the
QCD sum rules for vector mesons changes sizably in a nuclear medium. Under the basic
assumptions given above most of this change survives in the $N_c \to \infty$ limit.
We have shown that the in-medium four-quark condensate can be related to the
two-quark condensates in the large-$N_c$ limit. For $N_c=3$ the unknown size of the
four-quark condensate provides the major source of uncertainty on the condensate side.
On the hadronic side we have discussed various in-medium modifications. For $N_c=3$
none of them can be neglected {\it a priori}. It was a long-standing problem which
of these hadronic in-medium modifications (or which combination of them) corresponds
to the large in-medium change observed on the condensate side. We have found here
that the different hadronic effects show different scaling in powers of $1/N_c$.
In particular, in the large-$N_c$ limit only the coupling to resonance-hole loops
survives. Therefore this effect provides the hadronic counterpart of the large in-medium
change of the condensate side. It is
interesting to note that the hadronic effects which vanish in the large-$N_c$ limit
mainly cause a broadening of the spectrum and a (small) mass shift. On the other hand,
the coupling to resonance-hole excitations generates new peak structures. Hence the
large-$N_c$ analysis shows that the QCD sum rules do not point towards an in-medium 
mass shift or a peak broadening but indicate the appearance of one or several new peaks.
This is the qualitative result of our analysis. 

To become more quantitative we had to further specify the hadronic input. Here we were
guided by our phenomenological experience \cite{Peters:1998va,Post:2000qi,Post:2003hu}
that among the baryonic resonances it is the $D_{13}(1520)$ which provides most
of the in-medium change of $\rho$-mesons (at rest). In that spirit we have developed
an ansatz where the collective excitation of a $D_{13}$-resonance and a nucleon-hole 
couples to the $\rho$-meson as well as directly to the corresponding current. The
coupling constants were then determined from the QCD sum rules. In view of the 
approximations made (large-$N_c$ limit, neglecting all other resonances, 
non-relativistic model for $D_{13}$, neglecting the width of the $D_{13}$) 
we regard it as satisfying to obtain
reasonable values for the partial decay widths as discussed above.
In that context it is interesting to note that we have found that the obtained values 
for the coupling constants $f_\rho$ and $f_\gamma$ practically do not depend on the 
value for the $D_{13}$ mass. Therefore we do not expect that an analysis with a finite
$D_{13}$ decay width changes our results. Also if the mass of the $D_{13}$ in the
large-$N_c$ limit should significantly deviate from the physical one, we would not find 
a noticeable change --- provided that the scaling behavior given in Tab.~\ref{tab:tabnc}
remains untouched. We have also found that the in-medium change of the
threshold parameter is extremely small. In other words, the only in-medium change
that is seen in the QCD sum rule approach (at least in the large-$N_c$ limit) is
the collective excitation of the $D_{13}(1520)$ and a nucleon-hole.

\acknowledgments The authors thank U.~Mosel for discussions, encouragement and 
continuous support.
S.L.~also acknowledges stimulating discussions with M.~Lutz.

\bibliography{literature}
\bibliographystyle{apsrev}

\end{document}